 \documentclass[11pt]{article}
 \usepackage[top=1in,bottom=1in,left=1.5in,right=1.5in]{geometry}
 \usepackage{amsmath, amssymb, amsfonts}

  \newcommand{\hs}{\hspace*{\parindent}}

\begin{document}

 \title{Asymptotic Expansions for $\lambda_d$ of the Dimer and Monomer-Dimer Problems}

 \author
 {Paul Federbush \\
 Department of Mathematics\\
 University of Michigan \\
 Ann Arbor, MI 48109-1043 \\
 (\texttt{pfed@umich.edu})}
 
  \date{\today}
 \maketitle

 \begin{abstract}
In the past few years we have derived asymptotic expansions for $\lambda_d$ of the dimer problem and $\lambda_d(p)$ of the monomer-dimer problem. The many expansions so far computed are collected herein. We shine a light on results in two dimensions inspired by the work of M. E. Fisher.  Much of the work reported here was joint with Shmuel Friedland.
 \end{abstract}

We lament the rather two-dimensional nature of much of the current research on the dimer problem, and work in general $d$-dimensions. Given a unit rectangular lattice of volume $V$ the dimer problem loosely speaking is to count the number of different ways dimers (dominoes) may be laid down on the lattice, without overlapping, to completely cover it. Each dimer covers two nearest neighbor vertices. It is known that the number of such coverings is roughly exponential in the volume $\sim e^{\lambda_d V}$ for some constant $\lambda_d$ as $V$ goes to infinity. So it is desired to find $\lambda_d$. If one considers coverings that cover a fraction $p$ of the vertices one arrives at a constant $\lambda_d (p)$ instead. One has
\begin{equation}
\lambda_d(1) = \lambda_d.
\end{equation}

Over forty years ago the results of Hammersley and his collaborators, \cite{Ha1, Ha2, Ha3, Ha4} 
derived rigorous lower bounds for $\lambda_d$ and $\lambda_d (p)$ that may be said to presage the result

\begin{equation} \label{eq:2}
\lambda_d \sim \frac{1}{2} \ln (2d) - \frac{1}{2}.
\end{equation}
The rigorous bounds

 \begin{equation}\label{mincin}
 \frac{1}{2}\log (2d)-\frac{1}{2}\le \lambda_d \le \frac{1}{2}\log (2d)-\frac{1}{2}+\frac{\log(4 \pi d)}{4d}+\frac{1}{48} \frac{1}{d^2}
 \end{equation}
 were shown by Minc \cite{Min80}. Friedland, Kropp, Lundow, Markstr\"om, and Peled \cite{FKLM08, FP05} showed similarly for the monomer-dimer problem that
 
 \begin{align}\label{lpmdczd1}
& \frac{1}{2} \left(p\ln (2d)  -p\ln (p) - 2(1-p)\ln(1-p) -p \right)\le  \lambda_d(p)\le \nonumber  \\
  & \frac{1}{2} \left(p\ln (2d)  -p\ln (p) - 2(1-p)\ln(1-p) -p \right)+ p\left(\frac{\log(4\pi d)}{4d}-\frac{1}{48d^2}\right).
 \end{align}

In \cite{Fed09} I extended the result in \eqref{eq:2} to a full asymptotic expansion

 \begin{equation}\label{asexplam1}
 \lambda_d(1)\sim\frac{1}{2}\log (2d)-\frac{1}{2} +\sum_{k=1}^\infty \frac{c_k}{d^k}.
 \end{equation}
where the first three $c_k$'s have been numerically computed, \cite{Fed08A}
\begin{equation}
\lambda_d \sim \frac{1}{2} \ln (2d) - \frac{1}{2} + \frac{1}{8} \frac{1}{d} + \frac{5}{96}\frac{1}{d^2} + \frac{5}{64} \frac{1}{d^3}.
\end{equation}
(All of the asymptotic expansions we present have been formally derived; they await a rigorous proof, but are certainly true.) \cite{Fed09} is a three page paper that contains all the basic ideas used to derive our expansions!

In \cite{FF11} Shmuel Friedland and the author treated the monomer-dimer problem in a similar way, arriving at

 \begin{equation}\label{eq:asexplamp}
 \lambda_d(p) \sim \frac{1}{2} \left(p\ln (2d)  -p\ln p - 2(1-p)\ln(1-p) -p \right)+\sum_{k=1}^\infty \frac{c_k(p)}{d^k},
 \end{equation}
computed through the $k=3$ term as

\begin{align} \label{eq:8}
\lambda_d(p) \sim & \frac{1}{2} ( p \ln(2d) - p \ln p - 2(1-p) \ln (1-p) - p) \nonumber \\
& +  \frac{1}{8} \frac{p^2}{d} + \frac{(2p^3 + 3p^4)}{96} \frac{1}{d^2} + \frac{(-5p^4 + 12p^5 + 8p^6)}{192} \frac{1}{d^3}.
\end{align}

Shmuel and I came to believe that $\lambda_d(p)$ is better studied as a power series in $p$. Rearranging the terms in \eqref{eq:8} (and adding in some other known \emph{parts} of further terms in \eqref{eq:asexplamp} than those appearing in \eqref{eq:8}) we get
\begin{equation} \label{eq:9}
\lambda_d(p) \sim \frac{1}{2} (p \ln (2d) - p \ln p - 2(1 - p) \ln (1-p) - p) + \sum_{k=2} a_k(d) p^k.
\end{equation}
where the $a_k(d)$ are known for $k \leq 6$:
\begin{align}
a_2(d) &= \frac{1}{8} \frac{1}{d} \\
a_3(d) & = \frac{1}{48} \frac{1}{d^2} \\
a_4(d) & = \frac{1}{32} \frac{1}{d^2} - \frac{5}{192} \frac{1}{d^3} \\
a_5 (d) & = \frac{1}{16} \frac{1}{d^3} - \frac{39}{640} \frac{1}{d^4} \\
a_6 (d) & = \frac{1}{24} \frac{1}{d^3} - \frac{1}{32} \frac{1}{d^4} - \frac{19}{1920} \frac{1}{d^5} .
\end{align}
In \cite{FedConv} I offer a rigorous proof that the sum in \eqref{eq:9} converges if $p$ is small enough, but not that it converges to the correct value. But in \cite{FF11} we have many tests to show that it does converge to the correct value. We conjecture the right side of \eqref{eq:9} actually converges to the left side for all physical values of $d$ and $p$, i.e. $d=1,2, \ldots$, $0 \leq p \leq 1$.

We turn to the special case $d=2$ where somewhat more is known. In particular one has an analytic expression for $\lambda_2$.

\begin{equation}
\lambda_2 = \frac{1}{\pi} \sum_{k=0}^\infty (-1)^k \frac{1}{(2k+1)^2} = 0.291560904 \ldots
\end{equation}
This was established by M. E. Fisher \cite{Fisher61} and Kasteleyn \cite{Kas61}, in interesting and clever papers. I noted that in \cite{Fisher61} M. E. Fisher indicated steps he took in an unsuccessful effort to find an analytic expression for $\lambda_2(p)$. This helped prod me to a similar unsuccessful effort. As a first step I computed the expansion in \eqref{eq:9} to one further term beyond those for the general dimensional case, \cite{FedPTerm},

\begin{align} \label{eq:16}
\lambda_2(p) \sim & \frac{1}{2} (p \ln (4) - p \ln p - 2(1-p) \ln (1-p) -p) + \frac{4}{2} \cdot \left( \frac{1}{2 \cdot 1} \left(\frac{p}{4}\right)^2 \right.  \nonumber \\
&\left. + \frac{1}{3 \cdot 2} \left(\frac{p}{4} \right)^3 + \frac{7}{4 \cdot 3} \left( \frac{p}{4} \right)^3 + \frac{41}{5 \cdot 4} \left( \frac{p}{4} \right)^5 + \frac{181}{6 \cdot 5} \left( \frac{p}{4} \right) ^6 + \frac{757}{7 \cdot 6} \left( \frac{p}{4} \right) ^7 \right)
\end{align}
Our effort to find an analytic expression for $\lambda_2(p)$ was centered primarily on finding a ``natural'' extrapolation of the sequence
\begin{equation}
1,7,41,181,757
\end{equation}
an effort upon which I spent an embarrassing amount of time, my quest for the holy grail. I now feel such a quest is a priori doomed. In \cite{FedLattices} expansions analogous to \eqref{eq:16} are found for the two dimensional triangular and hexagonal lattices.

We thank one of the referees of this paper for asking about the relation of the series in (16) to the
virial expansion of a gas of dimers. Given either series to a given power of p, the other is 
determined to the same power. This transition was implicitly used in [8]. See the remark there in
the paragraph following the paragraph containing eq.(30), the sentence beginning: "By a simple
device...". But we were then ignorant of how natural it would have been to frame this development
in terms of the virial expansion. We plan to present details in a short upcoming paper.

\end{document}